\documentclass[twocolumn,prl,showpacs]{revtex4}

\usepackage{graphicx}
\usepackage{rotating}
\usepackage{amsmath}
\usepackage{amsfonts}
\usepackage{amssymb}
\usepackage{enumerate}
\usepackage{longtable}
\usepackage{dcolumn}% Align table columns on decimal point
\usepackage{bm}

\newcommand{\be}{\begin{equation}}
\newcommand{\ee}{\end{equation}}
\newcommand{\bn}{\begin{eqnarray}}
\newcommand{\en}{\end{eqnarray}}

\begin{document}

\author {M.S. Laad}
\title{Ferromagnetism in the Extended Periodic Anderson Model Near 
Selective Mott Localization.}
\affiliation{Max-Planck-Institut f\"ur Physik Komplexer Systeme,
01187 Dresden, Germany
}

\date{\rm\today}

\begin{abstract}
Motivated by the experimental finding of Ferromagnetism (FM) near non-FL
quantum phase
transitions (QPT), I investigate FM in an Extended Periodic Anderson Model
(EPAM).  Using the Frobenius-Perron theorem, a FM {\it metallic} phase near
the selective Mott localization is rigorously found.  Interesting connections
with Tasaki's flat-band ferromagnetism, as well as with the well-known double
exchange (DE) mechanism are elucidated.  This finding is a higher dimensional
generalization of numerical work on the $D=1$ PAM, and constitutes a robust way
to FM metallic behavior in $D>1$.  Finally, we discuss a concrete manifestation
of our results to $UGe_{2}$.
\end{abstract}

\maketitle

Itinerant Ferromagnetism (FM) in solids is a long-standing topic of interest
in condensed matter physics.
Progress in understanding FM had to await the quantum revolution.  FM is
a quintessentially quantum phenomenon, and has spurred intense theoretical
efforts~\cite{[1]}:  the problem is subtle, mandating a careful treatment of
the interplay between kinetic energy gain (by delocalization) and potential
energy gain (by localization).  We recall that precisely the same interplay
underpins another core issue of correlation-driven metal-insulator transitions
in $d$-band compounds~\cite{[2]}.
The connections between the two phenomena, and the resulting set(s) of
conditions under which FM is possible in appropriate model Hamiltonians are of
enduring interest.
Ingenious schemes have been constructed~\cite{[lieb],[tasaki],[5]} in this
context.
Tasaki's review summarizes the extensive study conducted
over the years to find FM in {\it fermionic} models.
In {\it real} materials, orbital degeneracy is considered to be
{\it essential} for stable FM phases, where the so-called
{\it double exchange} (DE)
mechanism
operates~\cite{[zener]}.

  In contrast to $d$-band systems, FM in $f$-band materials has been scantily
addressed.  Most conclusions are obtained from numerical
studies~\cite{[bat],[sigrist]}, apart from Sigrist's FM in the $J>>t$ Kondo
lattice in $D=1$. In principle, the existence of narrow-band $f$
electrons and wide-band $c$-electrons hybridized with each other in an
interacting model, the Periodic Anderson Model (PAM), should {\it favor} FM,
given its similarity to multi-orbital models, where one can
easily escape Pauli exclusion restrictions, except close to half-filling.
Strong correlations in the narrow $f$-band could Mott-localize the
$f$-electrons, creating local moments, and, with a partially filled $c$-band
electrons interacting with $f$-moments, a DE-like mechanism
could give a FM metallic state.  Whether
FM can arise {\it without} rigorous Mott localization of the $f$ electrons is
more subtle.
Nor is the issue purely of academic interest~\cite{[montu],[geibel]}.  In
$f$-band systems,
sizable $f-c$ hybridization is always present, and the $f$-electrons cannot be
viewed as localized in a strict sense.

  Here, I address the possibility of unearthing a FM metallic {\it phase} in an
extended PAM (christened EPAM) containing a {\it finite} $f-f$ hopping, $t_{f}$, an {\it extended} $f-c$ hybridization, $V_{fc}$, and a direct $f-c$
interaction, $U_{fc}$, in addition to the usual PAM.  Hence, the $f$-electrons
are {\it never} rigorously localized.  Naively,
 one might associate this with a renormalized itinerant picture
at low energies.  However, as shown earlier~\cite{[msl1],[msl2]}, a peculiar
interference arising from interplay between $t_{f,c}, V_{fc}$ and $U_{fc}$
can destroy the heavy FL state.  Given that this nFL metal is an {\it unstable} fixed point, one, or more broken symmetry phases will pre-empt it at lower $T$.  How might this happen?  Is it possible to study this issue, as rigorously as 
possible? 

  I start with the Hamiltonian, $H=H_{0}+H_{1}$, with
\be
H_{0}=t_{f}\sum_{<i,j>,\sigma}f_{i\sigma}^{\dag}f_{j\sigma} + t_{c}\sum_{<i,j>,\sigma}c_{i\sigma}^{\dag}c_{j\sigma} + V_{fc}\sum_{<i,j>,\sigma}f_{i\sigma}^{\dag}c_{j\sigma}
\ee
and

\be
H_{1}=U_{ff}\sum_{i}n_{if\uparrow}n_{if\downarrow} + U_{fc}\sum_{i,\sigma,\sigma'}n_{if\sigma}n_{ic\sigma'} +(\epsilon_{f}-E_{F})\sum_{i}n_{if}
\ee

Following earlier
work~\cite{[msl1],[msl2]}, I take:

(i) $V_{fc1}=\sqrt{t_{c}t_{f}}$ and $U_{ff}=\infty$.

(ii) I choose $\epsilon_{f}$ to lie close to the bottom of the $c$-band to
begin with: this condition will be made more precise below.  It could be
relaxed later.  But the exact argument goes through for this specific case, in
what follows.

(iii) the {\it total} band-filling, $n=n_{c}+n_{f}$, is chosen such that
$\epsilon_{f}=E_{F}$.

  Under these conditions, in an earlier work, I have studied the selective
 Mott QCP in $f$-band systems, and applied it to attempt an understanding of the
 singular non-FL metallic behavior in $YbRh_{2}Si_{2}$.  Remarkably, it will
 turn out that exactly these same conditions (but now with the additional condition (ii)) will be required for
 {\it proving} that the ferromagnetic
metallic state is the {\it rigorous}, unique ground state of the EPAM for a
critical, {\it finite} density.

To diagonalize $H_{0}$, but with $U_{ff}=\infty$, I introduce the combinations
$a_{i\sigma}=uX_{if\sigma}+vc_{i\sigma}, b_{i\sigma}=vX_{if\sigma}-uc_{i\sigma}$, with $u^{2}=t_{f}/t,v=t_{c}/t$ and the $X_{if\sigma}=(1-n_{if,-\sigma})f_{i\sigma}$ are projected fermion
 operators satisfying $[X_{if\sigma},X_{jf\sigma'}^{\dag}]_{+}=\delta_{ij}\delta_{\sigma\sigma'}(1-n_{if,-\sigma})$.  Also, $t=(t_{c}+t_{f})$.  In terms of 
these, $H$ becomes the spin $S=1/2$ Falicov-Kimball model (FKM) with a 
{\it local} $a-b$ hybridization:

\be
H_{FKM}=t\sum_{<i,j>,\sigma}(a_{i\sigma}^{\dag}a_{j\sigma}+h.c) + U_{fc}\sum_{i}n_{ia}n_{ib}
\ee
and

\be
H_{hyb}= (\epsilon_{f}-E_{F})\sum_{i,\sigma}(n_{ia}+n_{ib}+a_{i\sigma}^{\dag}b_{i\sigma}+h.c)
\ee
and at $\epsilon_{f}=E_{F}$, I recover $H_{FKM}$ with $U_{ff}=\infty$.  In general, for arbitrary $V_{fc}$ and $\epsilon_{f}$, inter- and intra-site $a-b$
hybridization terms supplement $H_{FKM}$ above.  Obviously, both $H$ and
the effective $H_{FKM}$ commute with {\it both} $S_{total}^{z}$ and $S_{total}^{2}$.  However, pseudospin ($\tau^{+}=X_{f\sigma}^{\dag}c_{\sigma}, \tau^{-}=c_{\sigma}^{\dag}X_{f\sigma}, \tau^{z}=(n_{Xf\sigma}-n_{c\sigma})/2$) SU$(2)$
invariance is explicitly broken, as is manifest in the form of $H_{FKM}$.
We also notice that the term $U_{ff}n_{f\uparrow}n_{f\downarrow}$ transforms to
$U_{ff}[u^{4}n_{a\uparrow}n_{a\downarrow}+v^{4}n_{b\uparrow}n_{b\downarrow} +
u^{2}v^{2}\sum_{\sigma}n_{a\sigma}n_{b,-\sigma}]$ in the $a-b$ basis.  
This implies that {\it both}
the $a$- and $b$-states are singly occupied {\it and} the  ($a\sigma,b-\sigma$)
local configuration is forbidden.

At $V_{fc1}$ and $\epsilon_{f}=E_{F}$, the rigorous, {\it local} U$(1)$
invariance of $H_{FKM}$ implies, by Elitzur's theorem~\cite{[elitzur]}, that
the ``excitonic''
average, $\Delta_{ab}=\langle(a_{i\sigma}^{\dag}b_{i\sigma}+h.c)\rangle=0$.
This gives $\Delta_{cf}=\langle(X_{if\sigma}^{\dag}c_{i\sigma}+h.c)\rangle=-(2V_{fc1}/(t_{c}-t_{f}))\langle(n_{ic\sigma}-n_{if\sigma})\rangle$.  Also, using this
relation, $\langle(n_{ia\sigma}-n_{ib\sigma})\rangle=((t_{c}+t_{f})/2V_{fc1})\langle(X_{if\sigma}^{\dag}c_{i\sigma}+h.c)\rangle$.  Hence, in the original EPAM,
the selective Mott localization actually corresponds to a {\it finite}
``excitonic'' average, $\Delta_{cf}=\langle(X_{if\sigma}^{\dag}c_{i\sigma}+h.c)\rangle=(2V_{fc1}/t)\langle(n_{ia\sigma}-n_{ib\sigma})\rangle$, or a finite
$\langle\tau_{i}^{x}\rangle$.  These relations
also show that $\Delta_{cf}=0$ for $V_{fc}=0$ in the EPAM, in agreement with
{\it exact} arguments ruling out excitonic order in the FKM (spin is irrelevant for $\Delta_{cf}$).
This would imply electronic ferroelectricity~\cite{[batef]} (EFE).
Thus, the selective metal supports ferro-excitonic order, an intriguing
result.  Physically, the fact that the
$b$-fermions rigorously localize aids in formation of local electronic
polarization, $\langle {\bf P}_{i}\rangle=-e{\bf r}_{i}\langle(X_{if\sigma}^{\dag}c_{i\sigma}+h.c)\rangle \propto \langle \tau_{i}^{x}\rangle$, co-existing
with metallic behavior.

 How about FM?  First consider the EPAM at a chosen filling,
$n=n_{c}+n_{f}=n_{1}=n_{b}+1\ne 1$ per site:
i.e, the $b$-levels are
singly occupied (with $U_{ff}=\infty$), and there is a {\it single} additional electron in the dispersive $a$ band.
 In the original EPAM, one has a {\it partially} filled case.  The itinerant
 $c$-fermions fill up available Bloch states,
and, when $\epsilon_{f}=E_{F}$, the much narrower $f$-band states are
progressively filled.  Since $U_{ff}=\infty$, one has either singly-occupied
{\it or} empty $f$ states.  In the ``rotated'' FKM, the problem
(recollect that, with the $f$-band at the bottom
of the $c$-band, the effective FKM has the $b$-band also lying at the bottom of the $a$ band) is now reduced to that of
a band of localized $b$ fermions, interacting with the {\it single}, itinerant
 $a$-fermion via $U_{fc}$.  Moreover, the connectivity condition is always
satisfied on most ``simple'' lattice structures (see Tasaki's review for
a precise characterization), and, with nearest neighbor hopping, now solely of
the $a$-fermion, any state can be reached from any other by repeatedly applying $H$.  These are precisely the set of conditions under which the Frobenius
Perron theorem~\cite{[fp]} can be exploited for showing FM.  This is what
we now show.

Consider a large but finite-sized system of $N$ sites. In the $a-b$ basis,
for an arbitrary configuration of the {\it localized} $b$-fermions,
$|\alpha_{b}\rangle=|\sigma_{b1},\sigma_{b2},...,\sigma_{b,i-1},h_{i,\sigma},\sigma_{b,i+1},...,\sigma_{bN}\rangle$ with a single {\it electron}, now
necessarily in the $a$-orbital at site $i$,
we have $H_{ij}=\langle i|H_{FKM}|j\rangle=t$ for
$|i-j|=1$, and zero otherwise.
 The finite (spin-independent) $U_{fc}$ only adds diagonal entries in $H_{ij}$
in the $a-b$ basis, since, with immobile $b$-fermions, it acts like a one-body
scattering potential for the $a$-fermion.
  Since $|\alpha_{b}\rangle$ contains eigenstates
of $S_{total}^{z}$ but not $S_{total}^{2}$, the Hilbert space is separable into
sectors with fixed $S_{total}^{z}$.  In each $S_{total}^{z}$ sector, $H_{ij}$
has $z$ non-zero entries in each row (and column).  Clearly, the {\it connectivity} condition~\cite{[fp]} is satisfied in our case.  If $t$ is chosen positive,
the Frobenius-Perron (FP) theorem implies that the {\it nodeless} eigenvector
has positive eigenvalue, $E_{max}=zt$, and is non-degenerate, which is the
ground state (GS).  Since there is no other orthogonal eigenstate having
{\it all} co-efficients positive, this GS is unique.  This state clearly has
$S_{total}^{z}=S_{max}$, and, since $[H,S_{total}^{2}]=0$, the state with
$S_{total}^{2}=S_{max}(S_{max}+1)$ turns out to have the same energy in each
$S_{total}^{z}$-sector.  Since the (nodeless) state with $E_{max}=zt$ is the
unique GS in each $S_{total}^{z}$-sector, it must have the same
$S_{total}^{2}=S_{max}(S_{max}+1)$.  The GS is thus {\it ferromagnetic}.
Moreover, with arbitrary band-filling, $n=n_{a}+n_{b}<1$, it is also metallic.
  Notice that the proof only implies a FM GS in the sense that its spin is
an extensive quantity (scaling with $N$, the number of sites).  It does not
describe its structure, since the proof tells us nothing about the spin-spin
correlation function, e.g, $\chi_{ij}=\langle S_{i}^{z}S_{j}^{z}\rangle$.

  This FM is a special case of Tasaki's flat-band FM
in the EPAM at the selective Mott point.  It can also be interpreted in the
``double exchange''(DE) sense: since $U_{ff}=\infty$, the $b$-{\it spin}
configuration is composed of a large number of degenerate, arbitrary spin
configurations.  Since the hopping (necessarily of the $a$-hole) is
independent of spin, maximum kinetic energy gain occurs when the
``background'', localized $b$-spins are ferromagnetically polarized.
Also, with large $U_{ff}$, it is energetically unfavorable (actually, with
$U_{b}=U_{ff}(t_{f}/t)^{2}=\infty$ in $H_{FKM}$, this is forbidden) to have
the $a\uparrow,b\downarrow$ local configuration, again stabilizing the FMM
state.
Since $f$-electrons are never strictly localized,
(only $b_{\sigma}=(vX_{f\sigma}-uc_{\sigma})$ is ``localized'' (SMT)),
the analogy with DE is
only a formal mathematical one.  Thus, our proof is a demonstration of
{\it itinerant} FM in the EPAM.  Further, $n_{1}=n_{b}+1=n_{f}+n_{c}$ implies
 finite itinerant ($c$) electron density in the original
EPAM.

{\it This means that, under the conditions on the parameters as chosen
above, a ferromagnetic metallic state is the rigorous ground state of the
EPAM at finite electron density}.

  This is, to my knowledge, the {\it first}
rigorous demonstration of a FMM ground state in the {\it finite density} EPAM.  
  In previous work~\cite{[msl1],[msl2]}, I have shown how selective Mott localization drives non-FL behavior in the {\it uniform} phase of the EPAM.  Since
the $b$-fermions are rigorously localized for $V_{fc1},\epsilon_{f}=E_{F}$,
a finite $U_{fc}$ produces strong resonant scattering between locally degenerate
 configurations with $n_{b}=0,1$, i.e, produces singular $b$-fermion ``valence'' fluctuations.  In the DMFT~\cite{[msl2]}, this gives rise
to infra-red singularities in the local response functions, as a
consequence of the Anderson orthogonality catastrophe (AOC), and the symmetry
unbroken metallic state is a non-FL metal.  Notice, however, that the AOC will
occur in the symmetry-unbroken metallic phase of $H$ in {\it any} $D$ at this
special point, and so this conclusion should survive beyond DMFT.
In this selective-Mott (non-FL) state the local $b$-fermion and ``excitonic''
($a-b$) correlators show {\it branch cut} structure instead of a renormalized
pole structure, characteristic of {\it local} quantum critical
behavior:~\cite{[msl1],[msl2]}

\be
\rho_{bb}(\omega) \simeq \theta(\omega)|\omega|^{-(1-\eta)}
\ee
and

\be
\chi_{ab}"(\omega) \simeq \theta(\omega)|\omega|^{-(2\eta-\eta^{2})}
\ee
with $\eta=(\delta/\pi)^{2}$, and $\delta=$tan$^{-1}(U_{fc}/W)$ is the
unrenormalizable scattering phase shift in the impurity version of $H_{FKM}$.
Thus, the nFL state is characterized
by soft, interband excitons, and coupling of fermionic quasiparticles to these
soft excitons destroys FL behavior.
Interestingly, it turns out that, for $U_{ff}=\infty$, this nFL state is
unstable to a FM metal (FMM).

  How stable are these states to changes in $V_{fc},\epsilon_{f}$?  For
$V_{fc}\ne V_{fc1}$ or $\epsilon_{f}=E_{F}$, either an inter- or intra-site
hybridization term is added to $H_{FKM}$.  Diagonalizing the bilinear part of
the resulting $H$, two bands, one with dominant $a(c)$ character with large
width, and the second with dominant $b(f)$ character and narrow width, are
obtained.  Such models have been solved both within DMFT(QMC)~\cite{[voll]}
and large lattice-size QMC~\cite{[bat]} calculatons.  In both, a robust
FMM phase is obtained over a range of band-fillings.  Also interesting in this context is the work of Biermann {\it et al.}~\cite{[bier]}, where a non-FL
state is found to be unstable to a FMM state at lower $T$.  The nFL phase
is presumably caused by the AOC in their two-orbital model as well, but, in
contrast to the analytic selective-Mott picture~\cite{[msl2]}, it is hard to
extract $Z=0$ at low $T$ from DMFT(QMC).  Thus, the FMM phase found near the selective-Mott point in the EPAM seems to be stable as one moves away from
that point.  Interestingly, a ``DE-like'' picture, related to ours, is also
proposed by
Batista {\it et al.}~\cite{[bat]} in $D=1$.In all earlier studies, however,
the intimate link between the SMT in the EPAM and the FMM phase has never been
addressed.  The analysis here bares this connection, we believe, for the
first time.  Moreover, our findings are independent of dimensionality, and rely
only upon the existence of the SMT and $U_{ff}=\infty$.
Additionally, {\it both} the SMT and FMM support excitonic order.
The proof goes through for $U_{fc}=0$, but {\it not} for $t_{f}=0$.  At this
point, we have the $S=1/2$ FKM in the $f-c$ basis.  It is easy, from the above,
to infer that simultaneous satisfaction of
conditions (i)-(iii) forces only {\it one} electron in the $c$-band for the FP
theorem to work (now in the $f-c$ basis) in this case.  Neither does the model
exhibit excitonic order: the exact {\it local} U$(1)$ symmetry of the EPAM
at $t_{f}=0$ rigorously implies $\Delta_{cf}=0$, by Elitzur's theorem.
An analytic proof of FMM for large, but {\it finite} $U_{ff}$, remains elusive
for the {\it full} EPAM.  

  However, with $t_{f}=0$, and a {\it local} hybridization, $V$,
the analysis of Batista~\cite{[bat]} (where our $t_{f}$ is 
their $t_{a}$) can be exploited to show FM.  Diagonalizing $H_{0}$ in the 
EPAM gives two hybridized bands, $\alpha$ and $\beta$.  If $|V|<<t_{c}$, there 
are two subspaces in {\it each} band where the states have dominantly $f,(\phi)$ or $c,(\psi)$ character.  Batista {\it et al.} showed that, under these 
conditions, $H$ can be re-expressed, to a very good approximation, in terms of 
the $\psi,\phi$ fermionic combinations of the $\alpha_{k\sigma},\beta_{k\sigma}$ 
fermions in {\it real} space.  The term $U_{fc}\sum_{i,\sigma,\sigma'}n_{if\sigma}n_{ic\sigma'}$ transforms to $U_{fc}\sum_{r,\sigma,\sigma'}n_{r\sigma}^{\phi}n_{r'\sigma'}^{\psi}$.  It turns out then that $H(\psi,\phi)=H_{0}(\phi,\psi)+H_{1}(\phi,\psi)$ with $U=\infty$ and the
{\it narrow-band} $a$-level energy, $\epsilon_{a}$ at $E_{F}$ is closely
related to our $H_{FKM}$ at $V_{fc}=\sqrt{t_{f}t_{c}}, \epsilon_{f}=E_{F}$:

\be
H_{0}(\psi,\phi)\simeq t_{\psi}\sum_{r,r',\sigma}\psi_{r\sigma}^{\dag}\psi_{r'\sigma} + t_{\phi}\sum_{r,r',\sigma}\phi_{r\sigma}^{\dag}\phi_{r'\sigma} 
\ee 
and 

\be
H_{1}=U\sum_{r}n_{r,\uparrow}^{\phi}n_{r\downarrow}^{\phi} + U_{fc}\sum_{r,\sigma,\sigma'}n_{r\sigma}^{\phi}n_{r'\sigma'}^{\psi}
\ee
so that when $t_{a}(t_{f})=0$, the dispersionless $a,(f)$ level sits at 
$E_{F}$.  Once again, partial band occupancy of the $a,b$ orbitals implies 
partially filled $\psi,\phi$ bands.  When $U=\infty$ and the $\phi$ level
lies at the bottom of the $\psi$ band, the FP theorem should give a FM metallic 
ground state.  More transparently, this is understood 
as follows.  In any dimension, $D$, the $\phi$-fermion band will undergo 
Mott-Hubbard splitting for large $U$.  The $\psi$-fermions now hop in the 
background of {\it localized} moments associated with the $\phi$ fermions.
Clearly, at $U=\infty$, maximum kinetic energy gain occurs when the $\phi$
fermion spins are ferromagnetically aligned, {\it a la} double exchange.

  What could be the experimental fingerprints of such a FMM state near the
selective-Mott QCP?  Since the $b$-fermions are rigorously immobile at the
QCP, but acquire a finite, heavy mass away from it, de-Haas van Alphen (dHvA)
measurements should show erosion of the heavy ($b$-fermion) Fermi sheet as the
QCP is approached.  Since entering the FM ordered phase drastically reduces
 the carrier scattering rate, as seen for example, in DMFT work~\cite{[bier]},
the $a$-fermion FS sheet should reflect {\it coherent} quasiparticles,
manifesting in well-defined quantum oscillations in the dHvA frequencies as
function of applied magnetic field.  These oscillations should become ill
defined in the incoherent nFL regime, since the large damping
(Im$\Sigma_{a}(\omega)$) within DMFT~\cite{[msl2],[bier]} will overdamp the
dHvA frequencies.  Away from the SMT-QCP, appearance of the second, heavy
$b$-fermion FS sheet should lead to additional dHvA frequencies.  We emphasize
that only the $b_{\sigma}$ combination, rather than the $f$-electrons
themselves, are localized in the EPAM.  Within the local, DMFT picture of the
SMT, anomalous power-law responses should characterize the thermodynamic,
transport and magnetic responses in the critical region: these should evolve
smoothly into coherent low-energy forms as one moves away from the QCP.
Emergence of an FMM state is then tied to having $\epsilon_{f}\simeq E_{F}$.
We are not able to say whether this is a generic requirement for FMM in the 
EPAM(PAM), but our result suggests that it might be the case.

  Whether the soft {\it interband} excitonic modes found at the SMT can act as
an electronic ``glue'' for unconventional multi-band SC is an interesting issue.
Since the soft triplet excitonic modes are generated by the orthogonality 
catastrophe, and so are not directly related to FM, they can even 
arise {\it within} the FM phase~\cite{[montu]}.  In fact, in
$UGe_{2}$, an additional line, $T_{x}(p)$, in the $T-p$ phase diagram is
known to exist.  Jumps in thermodynamic responses and dHvA frequencies across
$T_{x}(p)$ is suggestive of a kind of selective localization occuring across
that curve.  Interestingly, spin-triplet SC seems to be centered around the
QPT associated with $T_{x}(p)=0$, rather than with
$T_{c}^{FM}=0$~\cite{[montu]}.  In light of our analysis, it is tempting to
associate the unconventional, multi-band, spin-triplet SC in $UGe_{2}$ with
emergence of possible soft, spin-triplet excitonic modes around a hypotheisized
selective localization across $T_{x}(p)$.
An explicit calculation is out of scope of this work, however.

  In conclusion, I have studied the interesting issue of having an itinerant
FMM phase near the selective-Mott transition in an extended, periodic Anderson
model.  Using the FP theorem, I rigorously show, for $U_{ff}=\infty$ and
partially filled $c,f$ (or $a,b$) bands, that the FM state is the unique
{\it ground state}.
  Remarkably, the same feature, i.e, dispersionless $b$-fermion states,
which facilitates singular, non-FL dynamics in the uniform metallic phase,
also turns out to be crucial to prove FM in the EPAM.  Moreover, this partial
localization is also characterized by {\it soft}, interband excitonic modes
in the non-FL phase as a consequence of the Anderson orthogonality catastrophe.
In a FM {\it ordered} background, such soft interband spin-triplet excitons
could induce unconventional, spin-triplet SC.

{\bf Acknowledgements}  I thank Montu Saxena for a discussion on $UGe_{2}$ and
the MPI-PKS, Dresden for financial support.


\begin{thebibliography}{17}

\bibitem{[1]} C. Herring, in ``{\it Magnetism}'', Vol IV, eds. G. Rado and H. Suhl (Academic, NY, 1966).

\bibitem{[2]} J. Hubbard, Proc. Roy. Soc. (London) {\bf A276}, 238 (1963).

\bibitem{[lieb]} E. H. Lieb, Phys. Rev. Lett. {\bf 62}, 1927 (1989).

\bibitem{[tasaki]} H. Tasaki, Phys. Rev. Lett. {\bf 69}, 1608 (1992); ibid
J. Stat. Phys. {\bf 84}, 535 (1996); ibid Prog. Theo. Phys. {\bf 99}, 4, 489 
(1998), and references therein.

\bibitem{[5]} E. M\"uller-Hartmann, J. Low. Temp. Phys. {\bf 99}, 349 (1995).

\bibitem{[zener]} C. Zener, Phys. Rev. {\bf 82}, 403 (1951).

\bibitem{[bat]} C. Batista, J. Bonca, and J. Gubernatis, Phys. Rev. B{\bf 68},
214430 (2003).

\bibitem{[sigrist]} H. Tsunetsugu, M. Sigrist, and K. Ueda, Rev. Mod. Phys. {\bf 69}, 809 (1997), and references therein.

\bibitem{[montu]} S. S. Saxena, {\it et al.}, Nature {\bf 604}, 587 (2000).
K. Sandeman, G.Lonzarich and A. Schofield, Phys. Rev. Lett. {\bf 90}, 167005
(2003) and Ref.[11] therein, where dHvA data for $UGe_{2}$ show a change in the character of the Fermi surface across $T_{x}(p)$, indicating selective localization of a subset of carriers.

\bibitem{[geibel]} P. Gegenwart {\it et al.}, Phys. Rev. Lett. {\bf 94},
076402 (2005).

\bibitem{[msl1]} M. S. Laad, cond-mat/0805.2115, submitted tp Phys. Rev. Lett.

\bibitem{[msl2]} M. S. Laad, arXiv:0811.3767, submitted to Phys. Rev. Lett.

\bibitem{[elitzur]} S. Elitzur, Phys. Rev. D {\bf 12}, 3978 (1975).

\bibitem{[batef]} C. Batista, Phys. Rev. Lett. {\bf 89}, 166403 (2002).

\bibitem{[fp]} R.A. Horn and C.R. Johnson, Matrix Analysis, Cambridge University Press, 1990 (chapter 8).

\bibitem{[voll]} U. Yu, K. Byczuk, and D. Vollhardt, cond-mat/0808.0079, and
references therein.

\bibitem{[bier]} S. Biermann, L. de'Medici and A. Georges, Phys. Rev. Lett. {\bf 95}, 206401 (2005).






\end{thebibliography}
\end{document}